\documentclass[cits]{PoS}

\title{Evolution of magnetic fields in galaxies in the frame of hierarchical structure formation cosmology: future tests with the SKA}

\ShortTitle{Evolution of magnetic fields in galaxies: future tests
with the SKA}

\author{\speaker{Tigran Arshakian}%
, Rainer Beck, Marita Krause\\
        Max-Plank-Institut fuer Radioastronomie\\
        E-mail: \email{tarshakian;rbeck;mkrause@mpifr-bonn.mpg.de}}

\author{Dmitry Sokoloff\\
        Department of Physics, Moscow State University, Russia\\
        E-mail: \email{sokoloff@dds.srcc.msu.su}}

\author{Rodion~Stepanov\\
        Institute of Continuous Media Mechanics, Korolyov str.~1,
        614061 Perm, Russia\\
        E-mail: \email{rodion@icmm.ru}}

\abstract{Results from simulations of hierarchical structure formation cosmology
provide a tool to develop an evolutionary model of regular magnetic
fields coupled to galaxy formation and evolution. We use the dynamo
theory to derive the timescales of amplification and ordering of
magnetic fields in disk and puffy galaxies. Galaxies similar to the
Milky Way formed their disks at $z\approx10$ and regular fields of
$\mu$G strength and a few kpc coherence length were generated within
2~Gyr (at $z\approx3$), but field ordering up to the coherence scale
of the galaxy size took another 6~Gyr (at $z\approx0.5$). Giant
galaxies formed their disk already at $z\approx10$, allowing more
efficient dynamo generation of strong regular fields (with kpc
coherence length) already at $z\approx4$. Dwarf galaxies should have
hosted fully coherent fields at $z\approx1$. This evolutionary
scenario and number of predictions of the model can be tested by
measurements of polarized synchrotron emission and Faraday rotation
with the planned Square Kilometre Array. This model is used to simulate the
evolution of regular fields in disk galaxies and the polarized radio sky as
part of the Square Kilometer Array Design Studies\footnote{http://s-cubed.physics.ox.ac.uk/} (SKADS).}

\FullConference{Panoramic Radio Astronomy: Wide-field 1-2 GHz research on galaxy
evolution\\
                 June 2-5 2009\\
                 Groningen, the Netherlands}

\begin{document}

\section{Introduction}
The presence of regular magnetic fields in the disks of nearby spiral galaxies is inferred from observations of polarized synchrotron emission and Faraday rotation \cite{beck05}.
The amplitude and structure of magnetic fields is successfully reproduced by the mean-field dynamo theory which allows the dynamo theory to be applied for distant galaxies to explore the evolution of magnetic fields in distant galaxies.
We now have sufficient evidence that strong magnetic fields were
present in the early Universe \cite{kronberg08,seymour08} and that synchrotron emission should be detected with future radio telescopes such as the Square
Kilometre Array (SKA). The SKA will spectacularly increase the
sensitivity and angular resolution of radio observations and allow us
to observe an enormous number of distant galaxies at similar resolution to that achievable for nearby galaxies today.

Regular large-scale magnetic
fields can be generated and amplified by the mean-field galactic
dynamo in high-redshift galaxies, provided that a gaseous, rotating
disk already exists. The formation of disk galaxies and the epoch
of this formation are fundamental problems in astronomy. High
resolution numerical simulations of disk formation in galaxies
demonstrated that a dynamical disk could be formed at redshifts $z\sim
5-6$ and even higher \cite{governato07,mayer08}. A
more robust understanding of the history of magnetism in young galaxies
may help to solve fundamental cosmological questions about the
formation and evolution of galaxies.

We develop a simplified model for the evolution of magnetic fields in both
protogalactic halos and galaxies \cite{arshakian09} based on the recent numerical developments in the study of the formation and evolution of galaxies during the epoch of hierarchical structure formation.

\section{Evolution of magnetic fields in galaxies}

\begin{figure}[htb]
\begin{minipage}[t]{7cm}
\includegraphics[angle=-90,width=7cm]{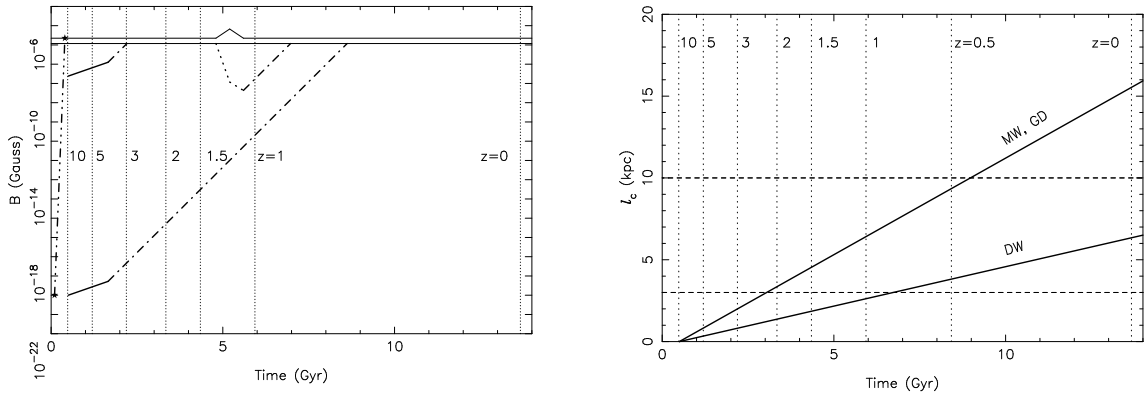}
\end{minipage}\hfill
\begin{minipage}[t]{7cm}
\includegraphics[angle=-90,width=7cm]{fig3.ps}
\end{minipage}
\caption{\emph{Left panel}. Evolution of magnetic fields in MW-type disk galaxies:
    magnetic-field strength versus cosmic epoch. Evolution of the small-scale magnetic
    field generated by the turbulent dynamo (thick dashed-dot-dot-dot line) and the
    large-scale magnetic field generated by the mean-field dynamo in quasi-spherical galaxies (thick solid line) or in thin-disk galaxies (thick dashed-dot-dashed). Dissipation of the field because of a major merger event is presented by a dotted line. The lower curve traces the evolution of regular magnetic fields generated by the pure large-scale dynamo mechanism (no amplification by
    the turbulent dynamo). The two horizontal thin solid lines represent the equipartition
    and equilibrium magnetic-field strengths
    The vertical thin dotted lines indicate redshifts from 0 to 10. \emph{Right panel}.
    Scale of ordering of regular magnetic fields with cosmological epoch.
    The evolution of the ordering of regular magnetic fields is shown for
    dwarf galaxies (DW; bottom line) and disk galaxies (MW and GD; top line).}
 \label{fig:mfe1}
\end{figure}

\emph{Three-phase model.} In the hierarchical formation scenario, we identify three main
phases of magnetic-field evolution in galaxies. In the first phase,
the seed magnetic fields of order $\simeq 10^{-18}$\,G
were generated in dark matter halos by the Biermann battery mechanism or Weibel instability \cite{lazar09}, well
before the formation of first massive stars at $z\sim20$ (Fig.~\ref{fig:mfe1}; left panel).

The second phase started at $z\sim20$ (cosmic time of $0.5$\,Gyr) when the
merging of dark-matter halos and virialization of infalling baryonic matter generated turbulence in the halo. According to \cite{wise07}, this epoch was dominated by
turbulence generated during the thermal virialization of halos.
There the turbulent (small-scale) dynamo could effectively amplify the seed
magnetic field of halos to the equilibrium level, $B_{\mathrm s}\sim 20~\mu$G, on timescales of few a hundreds million years.

\begin{figure}[t]
\parbox[s]{7.2cm}{
\includegraphics[angle=-90,width=7truecm]{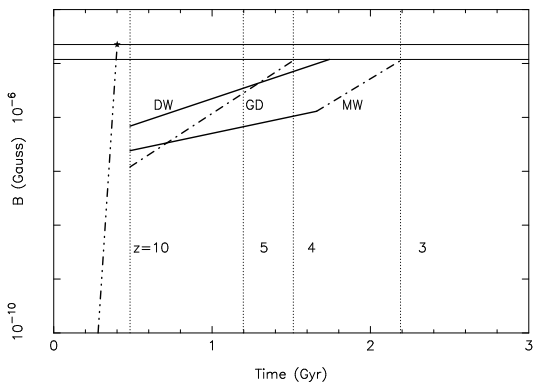} }
\hfill
\parbox[b]{7.5cm}{
\caption{Evolution in the magnetic-field strength of dwarf galaxies (DW), MW-type, and giant galaxies (GD). The meanings of different line types are the same as in Fig.~1. }
\label{fig:mfe2}
}
\end{figure}

In the third phase, the mean-field (large-scale) dynamo mechanism started
acting in the newly formed galaxies at $z\sim10$ (Fig.~\ref{fig:mfe1}; left panel). The first gas-rich
massive galaxies ($\gtrsim 10^9$ M$_{\odot}$) formed extended thin disks
($h/R\lesssim 0.1$, where $h$ and $R$ are the half-thickness and radius of the disk) at $z_{\rm disk}=10$ after major merger events. A weak, large-scale magnetic-field component was generated in the disk from small-scale magnetic fields of the halo, and then amplified to the equilibrium
level in the second phase. Then, the ``disk'' large-scale dynamo
amplified weak regular magnetic fields to the equipartition level ($\sim 10^{-5}$\,G) on times scales of $\sim1.5$\,Gyr (Figs.~\ref{fig:mfe1} and \ref{fig:mfe2}) and fully ordered regular magnetic fields on timescales of $\sim10$\,Gyr (Fig.~\ref{fig:mfe1}; right panel). If the formed disk was thick ($h/R \gtrsim 0.1$)
or the disk had not formed, the ``quasi-spherical'' mean-field
dynamo acting in puffy objects would be switched on. It would amplify the field
to the equilibrium level and order the magnetic fields on scales
of a few kiloparsecs and on timescales of
$\sim 1.3$\,Gyr and $\sim 6$\,Gyr ($h\approx R$),
respectively. If a thin disk in puffy objects had been formed at
later epochs, the ``disk'' mean-field dynamo would have dominated and amplified
the regular fields at the equilibrium level.

\begin{figure}
\center
\includegraphics[angle=0,width=15.3cm]{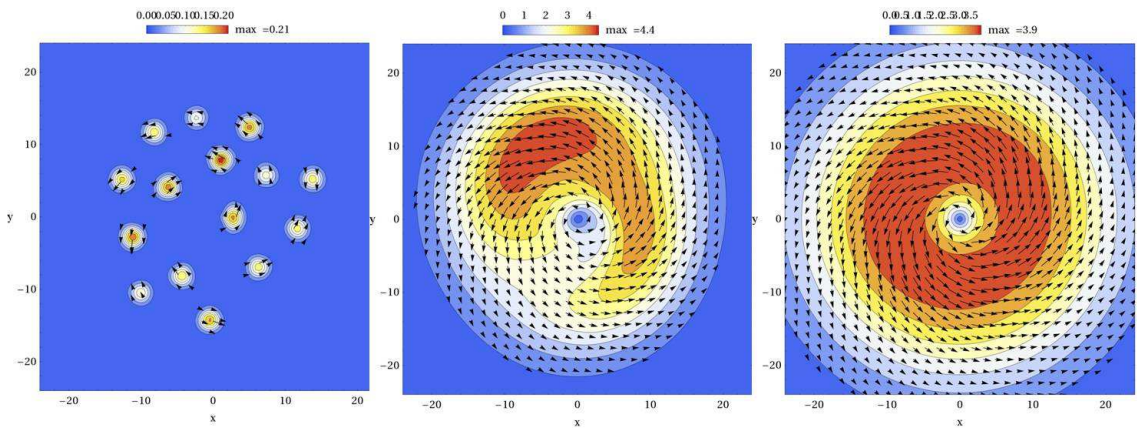}

    \caption{Simulations in the framework of the SKA design studies: the evolution of regular magnetic fields in the disk of a galaxy. The amplitude and ordering scale of the regular fields at the epoch of disk formation ($3\times10^{-7}$\,G and 1 kpc; left panel), after 5 Gyr ($\sim 2\times10^{-5}$\,G and 6 kpc; middle panel), and after 10 Gyr ($\sim 2\times10^{-5}$\,G and 12 kpc; right panel). }
 \label{fig:mfes}
\end{figure}

We simulated the evolution of magnetic fields for disk galaxies (Arshakian et al., in preparation) using the timescales of amplification and ordering of regular magnetic fields derived in \cite{arshakian09}. For a MW-type galaxy ($R=10$\,kpc), we started from seed fields generated in the newly formed disk at $z=10$. The initial size of the seed (1\,kpc), amplitude of its regular field (distributed normally around $3\times10^{-7}$\,G) and pitch angle (scattered within $\pm 20^{\circ}$) are used to simulate the evolution of regular fields to later epochs (Fig.~\ref{fig:mfes}). After 5\,Gyr the regular field is amplified to the equipartition level while the field is ordered at scales of 5 kpc. After 10 Gyr the filed is fully coherent (Fig.~\ref{fig:mfes}).  \\

We consider the evolution of magnetic fields in giant disk galaxies
(called GD hereafter; $R$(z=0)\,$=20$\,kpc), MW-type galaxies (10\,kpc), and
dwarf galaxies (called DW hereafter; $R$(z=0)\,$=3$\,kpc). We assume
that the seeds of turbulent magnetic fields of strength $\approx 10^{-18}$\,G existed
in the protogalaxies of present disk
galaxies, at $z \approx 35$. Virial turbulence could amplify
turbulent magnetic fields in merging dark-matter haloes via the
small-scale dynamo from $10^{-18}$\,G seed field strength to reach
the equipartition field strength of $2.2 \times 10^{-5}$\,G at
$z\simeq 11$ (0.4\,Gyr) within a short period of time, $\approx 3
\times 10^8$\,yr (see Figs.~\ref{fig:mfe1} and \ref{fig:mfe2}).

\emph{MW-like galaxies.} A sketch of the evolution of the magnetic
fields in isolated MW-type galaxies with $M_{\rm MW}\sim
10^{11}M_{\odot}$ is shown in Fig.~\ref{fig:mfe1} and zoomed in Fig.~\ref{fig:mfe2}.
At $z_{\rm disk} \approx 10$, the disk was formed and
evolved in isolation (no major merger) to the present time. The size
of the galaxy increased as $R_{10}(z=10)=R\,(1+z)^{-0.45}=3.4$\,kpc and
$h/R_{10}=0.14>0.1$, and the ``quasi-spherical'' mean-field dynamo
amplified the field until $z\approx 4$, at which time, $h/R$ became less
then 0.1. The ``disk'' mean-field dynamo then became significantly more important and
amplified the regular large-scale field within $\approx 1.5$\,Gyr,
and reached its equilibrium state at $z\approx3$
(Fig.~\ref{fig:mfe2}). At this epoch, the regular field was ordered
on a scale of a few kiloparsecs (Fig.~\ref{fig:mfe1}; right panel) and finally
reached a coherence scale similar to the size of the MW (10\,kpc) at
$z\approx 0.4$.

The earliest regular magnetic fields of equipartition strength were
generated in $\approx 1.4$~Gyr after disk formation
(Fig.~\ref{fig:mfe1}; left panel), while the ordering of magnetic fields on a length scale similar to that of a MW-type galaxy was complete after $\approx 9$~Gyr
(Fig.~\ref{fig:mfe1}; right panel). Hence, present-day, MW-type galaxies are
expected to host fully ordered regular fields.

\emph{Giant galaxies}. Late-type disk galaxies of a disk scale
length $\gtrsim~10$~kpc are rare. In these giant disk galaxies ($M_{\rm
GD}\sim 10^{12}M_{\odot}$), the ratio $h/R_{10}(z=10)=0.5/6.8<0.1$
implies that the ``disk'' mean-field dynamo has already been switched on at
$z\approx 10$ (Fig.~\ref{fig:mfe2}), had amplified the regular magnetic
field within only $\approx 1$~Gyr, and reached the equilibrium
state already at $z\approx 4$. However, as shown in
Fig.~\ref{fig:mfe1} (right panel), the regular magnetic fields can be ordered only
on scales of 15~kpc until the present time. Hence, the mean-field
dynamo cannot generate coherent magnetic fields over the size of the
disks of giant galaxies ($R\gtrsim15$\,kpc).

\emph{Dwarf galaxies}. We assume that, in low-mass halos, the
turbulent magnetic fields evolve in the same way as in their massive
counterparts. N-body/smoothed particle hydrodynamics simulations
demonstrated that low-mass galaxies did not at first form thin disks
because of the ionizing UV background radiation, which prevented
cooling of the warm galactic gas (Kaufmann et al. 2007). They were
more spheroidal, puffy systems ($h/R\approx 0.3$) of rotational
velocities $\approx 40$ km s$^{-1}$ and mass $M\approx 10^{10}$
M$_{\odot}$. In thick galaxies, the ``quasi-spherical'' dynamo could
generate and amplify the regular magnetic field to the equilibrium
field strength within $\approx 1.5$~Gyr (at $z \approx 3.5$;
Fig.~\ref{fig:mfe2}). At $z\approx2$, the UV background intensity
decreases, resulting in the formation of a
thin disk in a small galaxy that preserves the strength and
ordering of the existing regular field. Although the ordering
timescale of dwarf galaxies is longer (Fig.~\ref{fig:mfe1}; right panel), fully
coherent regular magnetic fields were generated at earlier epochs
($z\approx 1$) because of the smaller sizes of dwarf galaxies.


\section{Influence of star formation and mergers}
\emph{Star formation} can be triggered in isolated galaxies by gravitational
instability, in interacting galaxies by minor and
major mergers, by tidal forces, leading to the compression of the
gas, and by interactions of high velocity HI diffuse clouds. High SFR causes high velocity turbulence of the ionized gas, which in turn can suppress the mean-field dynamo in the thick disk if the turbulent velocity of
the gas is $\gtrsim 11 \,\,\mathrm{km \,\, sec}^{-1}$ (for MW-type galaxies) which in turn puts an upper limit of $\mathrm{SFR}\lesssim 20$
$M_{\odot}\,{\mathrm{yr}^{-1}}$ \cite{arshakian09} up to which the action of the
large-scale dynamo is possible.

\emph{Major mergers} were rare but could alter the morphology of a
spiral galaxy and destroy its regular magnetic field. After the
merger, there were two possibilities, firstly, to form a disk galaxy
with spheroidal or bulge component, or secondly, to form an
elliptical galaxy without a disk.
Disk galaxies, which survived after a gas-rich major merger, formed a
thin-disk component immediately after the merger event and needed
$\approx 1.5$ Gyr to amplify the regular magnetic field to the
equilibrium level by the mean-field dynamo (Fig.~\ref{fig:mfe1}; left panel) and
$\approx 8$ Gyr to generate a coherent magnetic field on the length scale of the galaxy size (Fig.~\ref{fig:mfe1}; right panel). \emph{Minor mergers} were more frequent and may also have altered
the morphology (spiral into elliptical, spiral to spheroid),
increased the size and thickness of the disk, and controlled the star
formation rate. Depending on the mass ratio of
galaxies and the number of minor mergers, the disk could either have been
preserved, forming a spheroidal component, or destroyed,
forming multiple spheroids or an elliptical galaxy.

\section{Future observational tests with the SKA}
Our analysis has strong implications on the expectations of the future observations of magnetic fields with the SKA.
The total magnetic field can be measured by the observed total power radio emission, corrected for the thermal fraction of a galaxy, while the regular magnetic field can be traced by polarized synchrotron emission and by Faraday rotation. The tight radio -- far-infrared correlation in galaxies implies that radio synchrotron emission is an excellent tracer of star formation in galaxies, at least to distances of $z\simeq3$ \cite{seymour08}. However, its application to even higher redshifts depends crucially on the existence of magnetic fields at the equipartition level with turbulent gas motions, as well as the effect of the cosmic microwave background (CMB) energy density which becomes stronger at high redshifts ($\sim (1+z)^{4}$).

The small-scale dynamo, with the help of virial turbulence, can amplify
turbulent fields to the level of equipartition with turbulent
energy density within $\simeq 3\times10^8$ years \cite{arshakian09}; strong
fields should therefore exist in all star-forming galaxies at
$z\simeq10$ (Fig.~\ref{fig:mfe1}) and the radio--far-infrared
correlation should be valid for $z\lesssim10$. However, the strong CMB radiation at high redshifts will suppress the non-thermal continuum emission of a galaxy by means of inverse Compton losses of cosmic-ray electrons suggesting that radio--far-infrared correlation should evolve with infrared/radio ratios increasing with redshift \cite{murphy09}.  Suppression of the non-thermal component at high redshift ($z>3$) would leave only radiation in thermal (free-free) regime thus limiting the depth to which the SKA can detect star-forming galaxies. Deviations from nominal IR/radio ratios at high-z will provide
a means for constraining the presence and strength of magnetic fields in young galaxies \cite{murphy09}.
The SKA and its pathfinder telescopes will investigate this relation in
more detail.

Number of predictions of the evolutionary model of magnetic fields can be tested with the SKA's measurements of polarized synchrotron emission and Faraday rotation of distant galaxies: (i) an anticorrelation at fixed redshift between galaxy size and the ratio between ordering scale and the galaxy size, (ii) weak regular fields (small Faraday rotation) in galaxies at $z\lesssim3$,
possibly associated with strong anisotropic fields (strong polarized
emission), would be signatures of major mergers, and (iii) undisturbed dwarf
galaxies should host fully coherent fields, giving rise to strong
Faraday rotation signals \cite{arshakian09}.

In the hierarchical merger formation model the massive galaxies with masses greater than the MW and high star-formation rate ($>100$\,M$_{\odot}$) form at $z\sim2-3$. On the other hand, observations of undisturbed massive galaxies at earlier epochs support the alternative idea of their formation by accretion of narrow streams of cold gas, as evident from cosmological simulations \cite{dekel09}. Polarization observations of massive distant galaxies with the SKA will be crucial to distinguish between different cosmological
scenarios of formation and evolution of galaxies.

\emph{Acknowledgments.} This work is supported by the EC Framework Program 6, Square Kilometre Array Design Study (SKADS) and the DFG-RFBR project under grant 08-02-92881.

\end{document}